\documentclass[twocolumn,showpacs,aps,prl,nofootinbib]{revtex4}
\usepackage{multirow}
\usepackage{amsmath}
\usepackage{epsfig}
\usepackage{subfigure}
\usepackage{float}
\usepackage{verbatim} 
\usepackage{booktabs}

\usepackage{axodraw4j}
\usepackage{pstricks}
\usepackage{color}

\newcommand{\ie}{{i.e.}}
\newcommand{\eg}{{e.g.}}

\newcommand{\ce}[1]{Eq.~(\ref{#1})}

\newcommand{\cf}[1]{{Fig.~\ref{#1}}}
\newcommand{\cfs}[1]{{Figs.~\ref{#1}}}
\newcommand{\ct}[1]{{Table~(\ref{#1})}}
\newcommand{\ctd}[2]{{Tables~(\ref{#1}) \& (\ref{#2})}}

\begin{document}

\bigskip
\title{Higgs production via vector-boson fusion at NNLO in QCD}

\author{Paolo Bolzoni$^{a}$, Fabio Maltoni$^{b}$, Sven-Olaf Moch$^{a}$, Marco Zaro$^{b}$}
\affiliation{
$^{a}$Deutsches Elektronen-Synchrotron, DESY\\
Platanenallee 6, D-15738 Zeuthen, Germany\\
$^{b}$Center for Particle Physics and Phenomenology (CP3), 
Universit\'e Catholique de Louvain, B-1348 Louvain-la-Neuve, Belgium}

\begin{abstract}
  We present the total cross sections at next-to-next-to-leading order (NNLO) in the
  strong coupling for Higgs production via weak boson fusion. 
  Our results are obtained via the structure function approach, which builds upon 
  the approximate, though very accurate, factorization of the QCD
  corrections between the two quark lines. The theoretical uncertainty on the
  total cross sections at the LHC from higher order corrections 
  and the parton distribution uncertainties are estimated at the $2$\% level each
  for a wide range of Higgs boson masses.
\end{abstract}


\maketitle

One of the main aims for the now-running LHC collider machine is to elucidate 
the mechanism of electro-weak symmetry breaking, and in particular to 
determine whether a Standard Model Higgs boson exists or not.

Among the various production mechanisms for the Higgs boson, vector-boson fusion (VBF) 
offers certainly one of the most promising and interesting signals~\cite{Cahn:1983ip,Kane:1984bb,Kleiss:1986xp}. 
The corresponding cross section is second in size outnumbered only by the gluon-gluon fusion process, 
it decreases rather mildly with the Higgs boson mass $m_H$ 
and it is proportional to the tree-level Higgs coupling to the vector-bosons, $W,Z$. 
Moreover, it provides such an  experimentally  clean
signature with the presence of at least two jets in the forward/backward
rapidity region that a rich variety of decay modes can be searched for, 
opening the access to the very difficult measurements of the Higgs couplings~\cite{Zeppenfeld:2000td}.
Given the importance and the experimental prospects for searches of VBF signals it is 
an urgent task to provide the corresponding theory predictions with the best possible precision 
including quantum corrections.

Higgs production in VBF is a pure electroweak process at leading order (LO), see \cf{fig:vbf}.
However, at a hadron collider the QCD radiative corrections are typically sizable, 
and they have first been computed for the total cross-section now almost two
decades ago in the so-called structure function approach~\cite{Han:1992hr}.
More recently, the differential cross-section at next-to-leading order (NLO) accuracy in QCD 
has become available~\cite{Figy:2003nv} 
along with its implementation in a Monte-Carlo event generator~\cite{Nason:2009ai}, 
and also the full set of combined NLO QCD and electroweak corrections are now known~\cite{Ciccolini:2007ec}. 
The typical accuracy of the current QCD predictions can be estimated in the $5-10\%$ range.

In this letter we briefly present the results of the computation 
of the dominant contributions to VBF at next-to-next-to-leading order (NNLO) accuracy in QCD, 
which give rise to dramatic reductions of the theoretical uncertainties.
To that end, we are using the structure function approach. 
As we will argue in the following, this method, although not truly exact at NNLO, 
includes the bulk of the radiative corrections so that the remaining contributions, 
which are both, parametrically small and kinematically suppressed, can be safely neglected. 

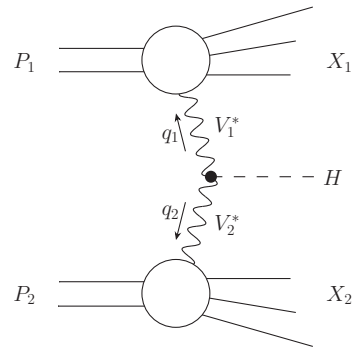
\begin{figure}[b!]
\centering
\scalebox{0.4}{
 \begin{picture}(260,322) (31,-15)
    \SetWidth{1.0}
    \SetColor{Black}
    \Line(33,267)(129,267)
    \Photon(144,258)(176,146){7.5}{6}
    \Photon(176,146)(144,34){7.5}{6}
    \Line[dash,dashsize=10](176,146)(273,146)
    \Line(32,24)(128,24)
    \Line(33,47)(129,47)
    \Line(33,245)(129,245)
    \Line(160,258)(256,274)
    \Line(160,34)(256,18)
    \Line(160,274)(272,306)
    \Line(160,18)(272,-14)
    \Line(155,50)(251,50)
    \Line(155,242)(251,242)
    \COval(143,36)(32,32)(0){Black}{White}
    \COval(143,256)(32,32)(0){Black}{White}
    \Text(292,146)[cc]{\huge{\Black{$H$}}}
    \Text(180,91)[lb]{\huge{\Black{$V^*_2$}}}
    \Text(180,184)[lb]{\huge{\Black{$V^*_1$}}}
    \Vertex(176,146){5.657}
    \Text(-10,246)[lb]{\huge{\Black{$P_1$}}}
    \Text(-10,26)[lb]{\huge{\Black{$P_2$}}}
    \Text(286,246)[lb]{\huge{\Black{$X_1$}}}
    \Text(286,26)[lb]{\huge{\Black{$X_2$}}}
    \Line[arrow,arrowpos=0,arrowlength=5,arrowwidth=2,arrowinset=0.35,flip](144,202)(152,170)
    \Text(130,178)[lb]{\huge{\Black{$q_1$}}}
    \Line[arrow,arrowpos=0,arrowlength=5,arrowwidth=2,arrowinset=0.35,flip](144,90)(152,122)
    \Text(130,106)[lb]{\huge{\Black{$q_2$}}}
  \end{picture}
  }
\caption{Higgs production via the VBF process.}
\label{fig:vbf}
\end{figure}

The first point to be addressed is to which extent VBF is a well-defined process by itself:
interference effects with other processes occur, possibly at higher orders in
the strong and/or electroweak coupling ($\alpha_s$ resp. $\alpha_{EW}$), 
whose size sets the target accuracy to which VBF as such can possibly be determined.
For example, VBF processes interfere already at LO 
with the so-called Higgs associated production, leading to a Higgs and two-jet final state, 
\ie, $pp \to HV^{(*)} \to Hjj$, 
an effect which is, however, at the per mil level~\cite{Ciccolini:2007ec}. 
At higher orders, interference with gluon-gluon fusion processes can occur
starting at $\alpha_{EW}^{2} \alpha_s^2$,
but they are also found to be
well below the percent level~\cite{Andersen:2006ag,Andersen:2007mp}.
It seems therefore conceivable that Higgs production through VBF 
can be defined 
in perturbation theory 
up to an ambiguity of not much better than 1\%, 
which sets the target precision for theoretical predictions.

The structure function approach is based on the observation that to a very good approximation 
the VBF process can be described as a double deep-inelastic scattering process (DIS), 
see \cf{fig:vbf}, 
where two (virtual) vector-bosons $V_i$ (independently) emitted from
the hadronic initial states fuse into a Higgs boson. 
This approximation builds on the absence (or smallness) of the QCD interference between 
the two inclusive final states $X_1$ and $X_2$. 
In that case the total cross section is given as a product 
of the matrix element $\mathcal M^{\mu\rho}$ for VBF, \ie, $V_1^\mu V_2^\rho \to H$, and 
of the DIS hadronic tensor $W_{\mu \nu}$, 
the latter being expressed in terms of the standard DIS structure functions $F_i(x,Q^2)$ with $i=1,2,3$:
\begin{widetext}
\begin{multline}
\label{eq:disapproach}
d\sigma=\frac{1}{2 S} 2 G_F^2 M^2_{V_1} M^2_{V_2} \frac{1}{\left(Q^2_1 +M^2_{V_1}\right)^2}
 \frac{1}{\left(Q^2_2 +M^2_{V_2}\right)^2}
 W_{\mu\nu} \left(x_1, Q^2_1\right) \mathcal M^{\mu\rho} \mathcal {M^*}^{\nu\sigma} W_{\rho\sigma}\left(x_2,Q^2_2\right)\times\\
 \times
 \frac{d^3 P_{X_1}}{\left(2\pi\right)^3 2 E_{X_1}}  \frac{d^3 P_{X_2}}{\left(2\pi\right)^3 2 E_{X_2}} ds_1 ds_2 \frac{d^3 P_H}{\left(2\pi\right)^3 2 E_H}
 \left(2\pi\right)^4 \delta^4\left( P_1+P_2-P_{X_1} -P_{X_2}-P_{H} \right) \, .
\end{multline}
\end{widetext}
Here $Q^2_i=-q_i^2$, $x_i=Q_i^2/(2P_i\cdot q_i)$ are the usual DIS variables, 
$s_i=(P_i+q_i)^2$ are the invariant masses of the $i$-th proton remnant, 
and $M_{V_i}$ denote the vector-boson masses, see \cf{fig:vbf}.
$G_F$ is Fermi's constant and at a given center-of-mass energy $\sqrt{S}$ of the collider 
the three-particle phase space is given by the second line in \ce{eq:disapproach}.

The factorization underlying \ce{eq:disapproach} does not hold exactly already at LO, 
because interference can occur either between identical final state quarks (\ie, $uu\to H uu$) or between
processes where either a $W$ or a $Z$ can be exchanged (\ie, $ud\to Hud$). 
However, at LO, they can be easily computed and have been included in our results.
On the other hand, simple arguments of kinematics 
(based on the behavior of the propagators in the matrix element~\cite{Dicus:1985zg}) 
show that such contributions are heavily suppressed and contribute to the total cross section well below the percent level, 
a fact that has been confirmed by complete calculation even through NLO~\cite{Ciccolini:2007ec}. 
Apart from these interference effects, the factorization of \ce{eq:disapproach} 
is still exact at NLO, so that the DIS structure functions at NLO~\cite{Bardeen:1978yd} can be employed.
This is due to color conservation: QCD corrections to the upper quark line, \cfs{diags}(a,b), are independent
from those of the lower line, 
\ie, Tr($t^a)=0$ for generators $t^a$ of the color SU$(N_c)$ gauge group.

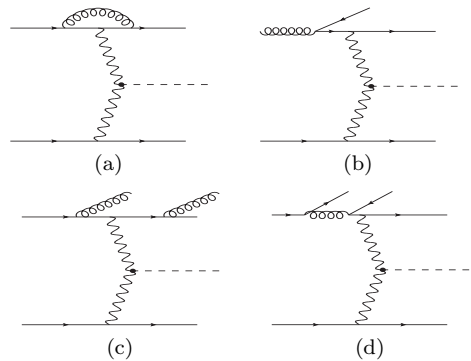
\begin{figure}[ht!]
\subfigure[]{\label{d1}
\resizebox{0.15\textwidth}{0.10\textwidth}{
\begin{picture}(290,268) (79,-23)
    \SetWidth{1.0}
    \SetColor{Black}
    \Line[arrow,arrowpos=0.5,arrowlength=5,arrowwidth=2,arrowinset=0.2](80,204)(208,204)
    \Line[arrow,arrowpos=0.5,arrowlength=5,arrowwidth=2,arrowinset=0.2](208,-20)(336,-20)
    \Line[arrow,arrowpos=0.5,arrowlength=5,arrowwidth=2,arrowinset=0.2](80,-20)(208,-20)
    \Line[arrow,arrowpos=0.5,arrowlength=5,arrowwidth=2,arrowinset=0.2](208,204)(336,204)
    \Photon(208,204)(240,92){7.5}{6}
    \Photon(208,-20)(240,92){7.5}{6}
    \Line[dash,dashsize=10](240,92)(368,92)
    \Vertex(242,92){4.472}
    \GluonArc[clock](208,184)(52,157.38,22.62){7.5}{8}
  \end{picture}
}
}  
\hspace*{.2cm}
\subfigure[]{\label{d2}
\resizebox{0.15\textwidth}{0.10\textwidth}{
 \begin{picture}(290,276) (79,-15)
    \SetWidth{1.0}
    \SetColor{Black}
    \Line[arrow,arrowpos=0.5,arrowlength=5,arrowwidth=2,arrowinset=0.2](160,212)(208,212)
    \Line[arrow,arrowpos=0.5,arrowlength=5,arrowwidth=2,arrowinset=0.2](208,-12)(336,-12)
    \Line[arrow,arrowpos=0.5,arrowlength=5,arrowwidth=2,arrowinset=0.2](80,-12)(208,-12)
    \Line[arrow,arrowpos=0.5,arrowlength=5,arrowwidth=2,arrowinset=0.2](208,212)(336,212)
    \Photon(208,212)(240,100){7.5}{6}
    \Photon(208,-12)(240,100){7.5}{6}
    \Line[dash,dashsize=10](240,100)(368,100)
    \Vertex(242,100){4.472}
    \Gluon(160,212)(80,212){7.5}{6}
    \Line[arrow,arrowpos=0.5,arrowlength=5,arrowwidth=2,arrowinset=0.2](240,260)(160,212)
  \end{picture}
}
}
\hspace*{.2cm}
\subfigure[]{\label{d3}
\resizebox{0.15\textwidth}{0.10\textwidth}{
  \begin{picture}(290,281) (79,-10)
    \SetWidth{1.0}
    \SetColor{Black}
    \Line[arrow,arrowpos=0.5,arrowlength=5,arrowwidth=2,arrowinset=0.2](80,217)(208,217)
    \Line[arrow,arrowpos=0.5,arrowlength=5,arrowwidth=2,arrowinset=0.2](208,-7)(336,-7)
    \Line[arrow,arrowpos=0.5,arrowlength=5,arrowwidth=2,arrowinset=0.2](80,-7)(208,-7)
    \Line[arrow,arrowpos=0.5,arrowlength=5,arrowwidth=2,arrowinset=0.2](208,217)(336,217)
    \Photon(208,217)(240,105){7.5}{6}
    \Photon(208,-7)(240,105){7.5}{6}
    \Line[dash,dashsize=10](240,105)(368,105)
    \Vertex(242,105){4.472}
    \Gluon(160,217)(240,265){7.5}{6}
    \Gluon(288,217)(368,265){7.5}{6}
  \end{picture}
  }
}
\hspace*{.2cm}
\subfigure[]{\label{d4}
\resizebox{0.15\textwidth}{0.10\textwidth}{
 \begin{picture}(290,276) (79,-15)
    \SetWidth{1.0}
    \SetColor{Black}
    \Line[arrow,arrowpos=0.5,arrowlength=5,arrowwidth=2,arrowinset=0.2](80,212)(128,212)
    \Line[arrow,arrowpos=0.5,arrowlength=5,arrowwidth=2,arrowinset=0.2](208,-12)(336,-12)
    \Line[arrow,arrowpos=0.5,arrowlength=5,arrowwidth=2,arrowinset=0.2](80,-12)(208,-12)
    \Line[arrow,arrowpos=0.5,arrowlength=5,arrowwidth=2,arrowinset=0.2](192,212)(336,212)
    \Photon(208,212)(240,100){7.5}{6}
    \Photon(208,-12)(240,100){7.5}{6}
    \Line[dash,dashsize=10](240,100)(368,100)
    \Vertex(242,100){4.472}
    \Line[arrow,arrowpos=0.5,arrowlength=5,arrowwidth=2,arrowinset=0.2](128,212)(192,260)
    \Line[arrow,arrowpos=0.5,arrowlength=5,arrowwidth=2,arrowinset=0.2,flip](192,212)(256,260)
    \Gluon(128,212)(192,212){7.5}{4}
  \end{picture}
  }
  }
\caption{Representative Feynman diagrams of processes included in the structure function approach.}
\label{diags}
\end{figure}

\begin{figure}[ht!]
\centering
\subfigure[]{\label{nd1}
\resizebox{0.15\textwidth}{0.10\textwidth}{
  \begin{picture}(240,230) (79,-61)
    \SetWidth{1.0}
    \SetColor{Black}
    \Line[arrow,arrowpos=0.5,arrowlength=5,arrowwidth=2,arrowinset=0.2](80,166)(208,166)
    \Line[arrow,arrowpos=0.5,arrowlength=5,arrowwidth=2,arrowinset=0.2](208,-58)(336,-58)
    \Line[arrow,arrowpos=0.5,arrowlength=5,arrowwidth=2,arrowinset=0.2](80,-58)(208,-58)
    \Line[arrow,arrowpos=0.5,arrowlength=5,arrowwidth=2,arrowinset=0.2](208,166)(336,166)
    \Photon(208,166)(240,54){7.5}{6}
    \Photon(208,-58)(240,54){7.5}{6}
    \Line[dash,dashsize=10](240,54)(368,54)
    \GluonArc(370,108)(200,163.25,236.25){7.5}{14}
    \Gluon(112,-58)(112,166){7.5}{12}
    \Vertex(242,54){6.325}
  \end{picture}
}
}
\hspace*{.2cm}
\subfigure[]{\label{nd2}
\resizebox{0.15\textwidth}{0.10\textwidth}{
   \begin{picture}(290,230) (79,-61)
    \SetWidth{1.0}
    \SetColor{Black}
    \Line[arrow,arrowpos=0.5,arrowlength=5,arrowwidth=2,arrowinset=0.2](80,166)(208,166)
    \Line[arrow,arrowpos=0.5,arrowlength=5,arrowwidth=2,arrowinset=0.2](208,-58)(336,-58)
    \Line[arrow,arrowpos=0.5,arrowlength=5,arrowwidth=2,arrowinset=0.2](80,-58)(208,-58)
    \Line[arrow,arrowpos=0.5,arrowlength=5,arrowwidth=2,arrowinset=0.2](208,166)(336,166)
    \Photon(208,166)(240,54){7.5}{6}
    \Photon(208,-58)(240,54){7.5}{6}
    \Line[dash,dashsize=10](240,54)(368,54)
    \Gluon(160,-58)(160,166){7.5}{12}
    \Gluon(285,166)(363,84){7.5}{7}
    \Vertex(242,54){6.325}
  \end{picture}
}
}
\hspace*{.2cm}
\subfigure[]{\label{tri}
\resizebox{0.15\textwidth}{0.10\textwidth}{
 \begin{picture}(277,242) (76,-36)
    \SetWidth{1.0}
    \SetColor{Black}
    \Line[arrow,arrowpos=0.5,arrowlength=5,arrowwidth=2,arrowinset=0.2](209,-33)(341,-33)
    \Line[arrow,arrowpos=0.5,arrowlength=5,arrowwidth=2,arrowinset=0.2](77,-33)(209,-33)
    \Gluon(77,198)(165,198){7.5}{6}
    \Gluon(253,198)(341,198){7.5}{6}
    \Line[arrow,arrowpos=0.5,arrowlength=5,arrowwidth=2,arrowinset=0.2](165,198)(253,198)
    \Line[arrow,arrowpos=0.5,arrowlength=5,arrowwidth=2,arrowinset=0.2](253,198)(208,130)
    \Line[arrow,arrowpos=0.5,arrowlength=5,arrowwidth=2,arrowinset=0.2](208,130)(165,198)
    \Photon(208,130)(231,77){7.5}{3}
    \Photon(231,77)(209,-33){7.5}{6}
    \Line[dash,dashsize=10](231,77)(352,77)
    \Vertex(231,77){7}
        \Text(195,170)[lb]{\Huge{\Black{$t,b$}}}
  \end{picture}
}
}
\hspace*{.2cm}
\subfigure[]{\label{box}
\resizebox{0.15\textwidth}{0.10\textwidth}{
 \begin{picture}(288,209) (65,-47)
    \SetWidth{1.0}
    \SetColor{Black}
    \Line[arrow,arrowpos=0.5,arrowlength=5,arrowwidth=2,arrowinset=0.2](154,-44)(341,-44)
    \Line[arrow,arrowpos=0.5,arrowlength=5,arrowwidth=2,arrowinset=0.2](66,-44)(154,-44)
    \Gluon(66,154)(154,154){7.5}{6}
    \Gluon(253,154)(341,154){7.5}{6}
    \Line[arrow,arrowpos=0.5,arrowlength=5,arrowwidth=2,arrowinset=0.2](154,154)(253,154)
    \Line[arrow,arrowpos=0.5,arrowlength=5,arrowwidth=2,arrowinset=0.2](253,154)(253,77)
    \Line[arrow,arrowpos=0.5,arrowlength=5,arrowwidth=2,arrowinset=0.2](154,77)(154,154)
    \Photon(154,77)(154,-44){7.5}{6}
    \Line[dash,dashsize=10](253,77)(352,49)
    \Line[arrow,arrowpos=0.5,arrowlength=5,arrowwidth=2,arrowinset=0.2](253,77)(154,77)
    \Text(165,110)[lb]{\Huge{\Black{$t,b$}}}
  \end{picture}
}
}
\caption{Representative Feynman diagrams of processes not included in the structure function approach.}
\label{nodiags}
\end{figure}
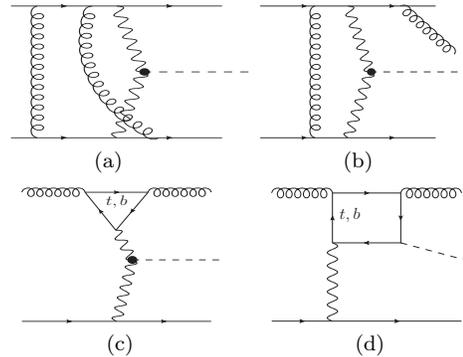

At NNLO due to the possibility of linking the upper and lower quark  lines
with two colored particles the factorization breaks down. However, it is easy
to see that such class of corrections, \cfs{nodiags}(a,b), are infrared and ultraviolet
finite, gauge invariant and suppressed~\cite{Figy:2007kv} both kinematically
and parametrically by a factor $1/N_c^2$.
The dominant contributions, \cfs{diags}(c,d) can therefore be included in  the
structure functions at NNLO~\cite{Kazakov:1990fu, Zijlstra:1992kj, Zijlstra:1992qd, Moch:1999eb}.
At order $\alpha_{EW}^3 \alpha_s^2$ another class of diagrams arises, \cfs{nodiags}(c,d),
which contributes significantly to associated Higgs production, $pp \to VH$, see~\cite{Kniehl:1990iva,Kniehl:1990zu}.  While a full computation of these diagrams  in VBF is in progress and will be presented elsewhere~\cite{bmmz2}, a first conservative estimate can be easily obtained in the $m_b \to 0, m_t \to \infty$ limit, where the triangle diagrams dominate.  In this limit, the contribution to the total cross section is well below the percent level and therefore can be neglected. 
At order $\alpha_{EW}^3 \alpha_s^2$, also di-jet amplitudes with a single quark line arise 
where the Higgs is emitted from a virtual weak boson loop. 
This class of diagrams is gauge invariant but not infrared safe and as it is not a VBF process, 
it is not included in our calculation. 
Its contribution to typical VBF final states with a Higgs and two jets 
has been found to be negligible \cite{Harlander:2008xn}.

\begin{figure}[t!]
\centering
\includegraphics[scale=0.6]{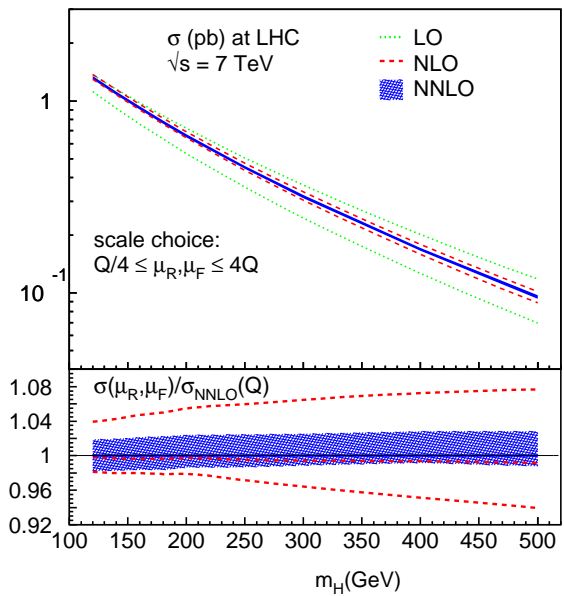}
\caption{
The total cross section at LO, NLO and NNLO 
as a function of $m_H$ 
for a $\sqrt{S}=7$ TeV LHC employing the MSTW PDF set~\cite{Martin:2009iq}.
The uncertainty bands are obtained by scale variation as explained in the text.}
\label{plot:mass}
\end{figure}

\begin{figure}[t!]
\centering
\includegraphics[scale=0.4]{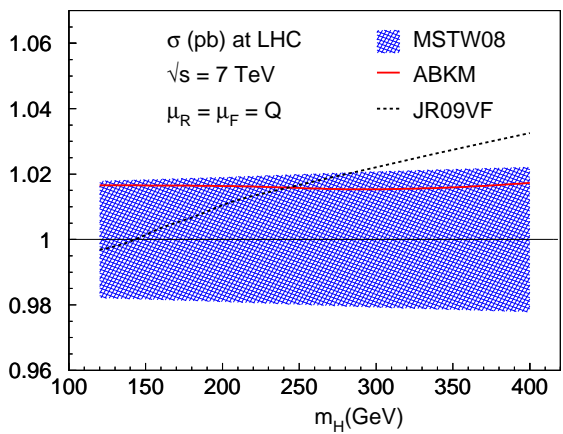}
\caption{The PDF uncertainty  of the total cross section at NNLO 
  as a function of $m_H$ at a $\sqrt{S}=7$ TeV LHC 
  for the 68\%  CL MSTW PDF set~\cite{Martin:2009iq}. 
  For ABKM~\cite{Alekhin:2009ni} and JR09VF~\cite{JimenezDelgado:2009tv} the ratio
  of the central value is plotted. 
}
\label{plot:pdf}
\end{figure}

We now turn to the discussion of the results. For the sake of illustration we
consider only a $\sqrt{S}=7$ TeV LHC, keeping in mind that the conclusions 
presented here are qualitatively the same 
for a $\sqrt{S}=14$ TeV LHC, and also for Tevatron, see~\cite{bmmz2}.
Our reference parton distribution functions (PDFs) set is MSTW~\cite{Martin:2009iq} and
the electroweak parameters ($G_F, M_Z,M_W, \sin^2 \theta_W$) are set to their 
respective PDG values~\cite{Amsler:2008zzb}.

\begin{center}
\begin{table}[tb!]
\begin{small}
\renewcommand{\arraystretch}{1.6}
\begin{center}
\begin{tabular}{|c|c|c|c|}
\toprule[0.08em]
\multicolumn{4}{|c|}{$\sqrt{S}=  7$ TeV}\\
\midrule[0.05em]
 Higgs mass &LO&NLO&NNLO\\[-2pt]
\midrule[0.05em]
120 &$  1.235^{+  0.131}_{-  0.116}$ &$  1.320^{+  0.054}_{-  0.022}$ &$  1.324^{+  0.025}_{-  0.024}$ \\
160 &$  0.857^{+  0.121}_{-  0.099}$ &$  0.915^{+  0.046}_{-  0.016}$ &$  0.918^{+  0.019}_{-  0.015}$ \\
200 &$  0.614^{+  0.106}_{-  0.082}$ &$  0.655^{+  0.038}_{-  0.012}$ &$  0.658^{+  0.015}_{-  0.010}$ \\
300 &$  0.295^{+  0.070}_{-  0.049}$ &$  0.314^{+  0.022}_{-  0.010}$ &$  0.316^{+  0.008}_{-  0.004}$ \\
400 &$  0.156^{+  0.045}_{-  0.030}$ &$  0.166^{+  0.013}_{-  0.007}$ &$  0.167^{+  0.005}_{-  0.001}$ \\
\bottomrule[0.08em]
\end{tabular}
\end{center}
\end{small}
\caption{
\label{tab:lhc}
Cross sections (pb) at a $\sqrt{S}=7$ TeV LHC with 
the uncertainty due to independent scale variations $\mu_R,\mu_F \in [Q/4,4Q]$ 
at LO, NLO and NNLO in QCD as obtained with the MSTW PDF sets~\cite{Martin:2009iq}.
}
\vspace*{-5mm}
\end{table}

\vspace*{-5mm}
\end{center}

\cf{plot:mass} presents the cross section as a function of the Higgs mass at LO, NLO and NNLO in QCD, 
together with the uncertainties coming from (uncalculated) higher orders. 
These are estimated by an independent variation of the factorization and renormalization scales in the range 
$\mu_R, \mu_F = \xi_{R,F} Q$ with $\xi_{R,F}\in [1/4,4]$, 
where $Q$ is the virtuality of the vector-boson probing the corresponding
structure function 
to which we apply a technical cutoff of $1$ GeV.
The lower inlay of \cf{plot:mass} zooms in on the relative variations 
normalized to the NNLO cross section at $\mu_R, \mu_F = Q$, 
so that the exceptionally good convergence of the perturbation series can be appreciated.
For NNLO this is at the 2\% level and in principle, could be pushed
even further within the structure function approach by incorporating the available hard corrections 
at order $\alpha_s^3$~\cite{Vermaseren:2005qc,Moch:2007rq,Moch:2008fj}.
Numbers for our best estimate, \ie, NNLO in QCD, are presented in \ct{tab:lhc}.

The most natural choice $\mu_R, \mu_F = \xi_{R,F} Q$ as a reference scale is also supported 
by kinematics arguments, \ie, the observation that the average gauge boson virtuality 
in VBF amounts only to $\langle Q \rangle \simeq 20$ GeV for a $\sqrt{S}=7$ TeV LHC.
Of course, other scale choices, \eg \ $\mu_R, \mu_F \in [m_H/4,4m_H]$, are equally valid.
However, they typically exhibit a much poorer convergence of the perturbative expansion and 
lead to sizable deviations in the lower order predictions, especially for heavy Higgs bosons 
(\eg\ a $7\%$ difference for $m_H = 400$ GeV at NLO).
Only at NNLO, both the central values and the uncertainty band for the latter choice agree 
within the 2\% level with those in \ct{tab:lhc}.
This clearly demonstrates the markedly improved scale stability of our NNLO predictions.

In \cf{plot:pdf} the dependence on the parton distributions and their errors is studied, 
which estimates the uncertainty of the total cross section due to
the non-perturbative parton dynamics inside the proton. 
To this aim we employ the MSTW 68\% confidence level PDF sets~\cite{Martin:2009iq} 
through NNLO and compare also with the
central predictions obtained with the other available PDF sets based on complete NNLO QCD predictions, 
\ie, ABKM~\cite{Alekhin:2009ni} and JR09VF~\cite{JimenezDelgado:2009tv}.
The results are consistent and show that an almost constant 2\% PDF uncertainty at NNLO
can be associated to the cross sections for a wide range of Higgs boson masses, 
which is slightly reduced compared to the NLO value of 2.5\%.
The PDFs are dominantly sampled at an average $\langle x \rangle \simeq 0.05$ at a $\sqrt{S}=7$ TeV LHC,
that is in a region where further non-perturbative corrections such as higher-twist
effects are negligible, see \eg, Ref.~\cite{Alekhin:2007fh}.

%
\begin{center}
\begin{table}[tb!]
\begin{small}
\renewcommand{\arraystretch}{1.6}
\begin{center}
\begin{tabular}{|c|c|c|}
\toprule[0.08em]
\multicolumn{3}{|c|}{$\sqrt{S}=  7$ TeV}\\
\midrule[0.05em]
 Higgs mass &NLO$^2$&NNLO$^2$\\[-2pt]
\midrule[0.05em]
120 &$  1.319^{+  0.054}_{-  0.020}$ &$  1.323^{+  0.024}_{-  0.022}$ \\
160 &$  0.914^{+  0.047}_{-  0.014}$ &$  0.918^{+  0.020}_{-  0.015}$ \\
200 &$  0.654^{+  0.039}_{-  0.011}$ &$  0.657^{+  0.016}_{-  0.011}$ \\
300 &$  0.313^{+  0.024}_{-  0.009}$ &$  0.315^{+  0.009}_{-  0.005}$ \\
400 &$  0.166^{+  0.014}_{-  0.006}$ &$  0.167^{+  0.005}_{-  0.002}$ \\
\bottomrule[0.08em]
\end{tabular}
\end{center}
\end{small}
\caption{
\label{tab:syst}
Cross sections (pb) at a $\sqrt{S}=7$ TeV LHC  
with the same parameters as \ct{tab:lhc}. 
The approximations NLO$^2$ (NNLO$^2$) employ the complete NLO (NNLO) expressions 
for the structure functions both on the upper and lower quark lines, cf. \cf{fig:vbf}.} 
\vspace*{-5mm}
\end{table}

\vspace*{-5mm}
\end{center}

Let us next turn to a discussion of the systematics of our approach.
In \ct{tab:syst} we present an alternative study of the perturbative series,
namely we consider the effects of ``improving'' the  $\alpha_s$ and $\alpha_s^2$ expansions 
of the cross section, in terms of expansions of the structure functions at NLO and NNLO which enter as a square. 
The NLO$^2$ are obtained by keeping the $\alpha_s$ terms in both structure functions, \ie,
including formally $\alpha_s^2$  terms in the cross section and using the NLO PDFs,
while the NNLO$^2$ results perform the same procedure at one higher order in $\alpha_s$.
A comparison of the numbers in \ctd{tab:lhc}{tab:syst} shows that the
different higher order approximations 
agree extremely well within the respective errors bands. 
Implicitly, our findings also demonstrate that the impact of the NNLO QCD corrections 
at the central value, \ie, $\mu_R, \mu_F = Q$, is relatively small, 
and in fact smaller than 1\%, as we have explicitly checked.
Altogether this gives further evidence that non-factorizable
contributions at NNLO which are not accounted for in the structure
function approach and can very conservatively be estimated to
amount to less than 10\% of the NNLO factorizable ones,
are completely negligible.

To summarize, we have presented the first computation of the VBF cross section at NNLO in QCD. 
The inclusion of higher order corrections stabilizes the results at the $2\%$ level 
against arbitrary variations of the renormalization
and factorization scales, indicating an extremely well-behaved perturbative expansion. 
PDF uncertaines are estimated at the 2\% level as well,
uniform over a wide range of Higgs boson masses. 
Our results motivate the calculation of the differential cross sections distributions at NNLO via an exclusive method such as that proposed in Ref.~\cite{Daleo:2009yj}.
Moreover, our approach can be used to provide cross section predictions to NNLO accuracy 
for any weak-boson fusion process leading to a weakly interacting $n$-particle
final state $X$, \ie, $V^*V'^*\to X$. Work in the latter direction is in progress. 

We thank Giuseppe Degrassi and Maria Ubiali for discussions. 
This work is partially supported by the Belgian Federal Office for 
Scientific, Technical and Cultural Affairs through Interuniversity Attraction Pole No. P6/11,
by the Deutsche Forschungsgemeinschaft in SFB/TR 9 and by the 
Helmholtz Gemeinschaft under contract VH-NG-105.


\end{document}